# *Machine Learning Prediction of Accurate Atomization Energies of Organic Molecules from Low-Fidelity Quantum Chemical Calculations*

Logan Ward[1,2], Ben Blaiszik[1,3], Ian Foster[1,2,3], Rajeev S. Assary[4,5], Badri Narayanan[5,6], Larry Curtiss[4,5]

1 Data Science and Learning Division, Argonne National Laboratory
2 Department of Computer Science, University of Chicago
3 Globus, University of Chicago
4 Joint Center for Energy Storage Research (JCESR), Argonne National Laboratory
5 Materials Science Division, Argonne National Laboratory
6 Department of Mechanical Engineering, University of Louisville, Louisville, KY

## Abstract
Recent studies illustrate how machine learning (ML) can be used to bypass a core challenge of molecular modeling: the tradeoff between accuracy and computational cost. Here, we assess multiple ML approaches for predicting the atomization energy of organic molecules. Our resulting models learn the difference between low-fidelity, B3LYP, and high-accuracy, G4MP2, atomization energies, and predict the G4MP2 atomization energy to 0.005 eV (mean absolute error) for molecules with less than 9 heavy atoms and 0.012 eV for a small set of molecules with between 10 and 14 heavy atoms. Our two best models, which have different accuracy/speed tradeoffs, enable the efficient prediction of G4MP2-level energies for large molecules and are available through a simple web interface.

## 1   Introduction

There are a large range of quantum chemical methods for calculation of molecular energies, with trade-offs between accuracy and computational cost governed by the approximations used to make the predictions computationally tractable.[1] One type of quantum chemical approach for accurate energy calculations is based on a composite technique in which a sequence of well-defined *ab initio* molecular orbital calculations is performed to determine the total energy of a given molecular species. Composite methods, such as G4MP2,[2] typically have a mean absolute error (MAE) accuracy of better than 1 kcal/mol (0.04 eV) for test sets of molecules with accurate data. However, at the current time the size of molecules to which these methods can be applied is limited by the availability of sufficient computational power. Density functional methods are much faster, but less accurate.[3] The widely used B3LYP DFT method[4] has a MAE of about 4 kcal/mol (0.2 eV), but is also limited in the size of the molecules that can be handled, due to time-consuming calculations.

Machine learning (ML) offers opportunities for bypassing the tradeoff between accuracy and computation cost. Recent publications have shown how ML can create fast models that directly link the inputs and outputs of an expensive calculation.[5–7] Further studies demonstrate the advantages of predicting the differences between low- and high-fidelity calculations (i.e., Δ-Learning),[8] or of using multiple levels of fidelity to train the same model.[9,10] It is also possible to link easily-computable properties to the outcomes of extensive calculations, as shown by how Seko et al. used calculated bulk moduli as inputs to a model that predicts melting temperature.[11] Neural networks offer more possibilities through the ability to train the same model on multiple properties (e.g., multi-task or transfer learning).[12] It is as yet unclear which of these many options yields optimal performance for different types of applications.

In this work, we focus on designing ML models to predict the atomization energy of molecules at G4MP2-level accuracy at a lower computational cost. In particular, we investigate how to tailor two of the best-performing methods for learning molecular properties to this task, namely: SchNet[13]—a deep convolutional neural network approach, and FCHL—a conventional machine learning approach.[14] We examine how to integrate information from low-cost B3LYP calculations into each method and find that both techniques predict atomization energies of molecules larger than our training set with errors below 0.04 eV using a Δ-learning approach. We have created a simple interface to allow others to use our best-performing models by publishing them on DLHub,[15] so that accurate atomization energies are readily accessible to the materials and chemistry community at large.

## 2 Methods

We first describe the datasets and ML approaches used in this work.

### 2.1 Datasets

We used the QM9-G4MP2 dataset described by Curtiss et al.[16] as a starting point for our model. This dataset contains the B3LYP- and G4MP2-computed properties for 133,296 molecules, each with from one to nine heavy atoms (C, F, N, O). The B3LYP data and the geometries for the molecules are from the QM9 dataset of Ramakrishnan et al.[17] We selected 130,258 of the 133,296 molecules for use in our study, omitting those whose bonding connectivity changed on relaxation, as identified by Ramakrishnan et al.. We randomly selected 10% of the remaining QM9-G4MP2 as the hold-out set to be used to evaluate model performance, but never used in model training or hyperparameter optimization, yielding a 117,232-molecule training dataset, QM9-G4MP2-train, and a 13,026-molecule hold-out set, QM9-G4MP2-holdout.

We also make use of a separate dataset, G4MP2-heavy, of G4MP2 energies for 66 molecules with between 10 and 14 heavy atoms, calculated previously as part of a study on bio-oil derived molecules. We used these molecules to evaluate the ability of our models to predict the properties of molecules larger than those in QM9-G4MP2, although we note this is a relatively small set and a set including more larger molecules is needed to more accurately evaluate our models.

The identities, molecular coordinates, and computed properties of the molecules in G4MP2-heavy and the exact train/test splits used for QM9-G4MP2-train are available in full on the Materials Data Facility[18,19] and GitHub.[20]

### 2.2 Machine Learning Approaches

We employ two ML strategies: the continuous-filter convolution neural networks of Schütt et al. (SchNet), and the alchemical and structural distribution approach of Faber et al. (FCHL).

#### 2.2.1 SchNet

We selected SchNet as a deep learning approach for predicting the properties of molecules, given its best-in-class performance on predicting atomization energies at the time this study began.[13] SchNet takes the atomic numbers and positions of each atom in a molecule as inputs. First, each atomic number is mapped to a vector (the "embedding") to generate the initial representation for each atom. The interaction layers of SchNet update these representations based on distances to, and representations of, nearby neighbors. The representation produced at the end of the interaction layers is then fed into a multi-layer, dense neural network to produce the contribution of each atom. The atomic contributions are then aggregated (e.g., via summation) to generate the molecular property.

We use the open-source implementation of SchNet available in SchNetPack[21] and the recommended hyperparameters defined in Ref. [21]. With these hyperparameters, a SchNet model has millions of trainable parameters, including the embeddings for each element, parameters for how distance relates to updated representation, and other model components. As with most deep neural networks, SchNet is trained by iteratively adjusting each parameter based on the gradient of the error with respect to each parameter, as computed via backpropagation. We employ the optimizer (Adam) and learning rate schedule specified in Ref. [21].

### 2.2.2 FCHL

The FCHL method uses Kernel-Ridge Regression (KRR) to learn molecular properties of an atomistic system from *M*-body representations of the local chemical environment.[14] The core of the FCHL method is an approach for measuring the similarity of the local environments of two atoms. Each atom is described using a series of *M*-body expansions, which are modeled as weighted sums of Gaussians over sets of the neighbors. The similarity of two atomic environments is computed as an integral over the squared difference between each of these distributions. The similarity of two molecules is defined as a sum of the similarities between each atom in each molecule, which leads to improved accuracy on training sets with diverse molecular sizes over similarity metrics that consider the molecule as the fundamental unit.[22] We use the recommended hyperparameters for this method, and the open-source QML library[23] and Scikit-learn[24] to fit FCHL models.

## 3   Results and Discussion

Our goal is to develop a model that predicts G4MP2 energies for organic molecules with accuracies comparable to G4MP2 computations. (G4MP2-computed enthalpies of formation, obtained from atomization energies of organic molecules, have a MAE of 0.77 kcal/mol when compared to accurate experimental values[1,16].) That is, we want a model that when trained on a set of (molecule description, G4MP2 energy) pairs (the training set) can achieve high accuracies when used to predict G4MP2 energies for other molecules for which only the description is provided (the test set).

In our work, we compared the performance of different modeling strategies, validated the ability of each strategy to compute the energies of molecules larger than the training set, and assessed the degree to which knowledge of a molecule's equilibrium structure effects accuracy. We explore each topic in turn.

### 3.1   Modifying SchNet to Incorporate Information from Low-Fidelity Calculations

The SchNet neural network architecture permits several routes to augmenting our predictions of G4MP2 energies with results from other calculations. We implemented five strategies in all to produce five models in addition to the baseline SchNet. Three require only a molecular structure as input:

> *SchNet Transfer:* Transfer learning in neural networks is often accomplished by using the weights learned in a related problem as a starting point in training a new model. We used the weights from a model trained on B3LYP atomization energies as a starting point for our model.

*SchNet Multitask*: Training a network on several related outputs is thought to cause models to learn more-generalizable representations.[25] We explore this strategy by concurrently training a SchNet model on B3LYP- and G4MP2-computed atomization energies and the B3LYP-computed HOMO, LUMO, and Zero Point Energy.

*SchNet Stacked*: Stacking in ML is the technique of using one model's output as an input to another. We use the atomic contributions to the total B3LYP energies as inputs to the output layer of SchNet, creating what is effectively a Δ-learning model (see SchNet Delta below) that infers the difference between B3LYP and G4MP2 energies, rather than the B3LYP energy directly.

The other two models use molecular/atomic properties computed with B3LYP as model inputs:

*SchNet Delta*: As introduced by Ramakrishnan et al., Δ-Learning models learn the difference between different calculations.[8] We train a model that learns the difference between B3LYP and G4MP2 energies.

*SchNet Charges*: We use the partial charges from B3LYP as features in the SchNet embedding layer, which originally contains only features related to the atomic element. (We also experimented with using partial charges in the output layers rather than the embedding layer, but did not find notable improvements in performance.)

All source code needed to create models using these approaches is available in a GitHub repository that includes scripts with the hyperparameter choices for our models, results showing that we replicate previous literature, and the exact versions of SchNetPack used in our study.[20] The modifications we made to SchNetPack will be contributed to the main repository after submission of this paper.

We first tested each model by performing a standard cross-validation test. Each model was trained by using identical subsets of QM9-G4MP2-train with sizes ranging from 1000 to 117,232 entries, all of which are available from MDF.[19] We trained each model until the learning rate decayed to $10^{-6}$ (as suggested by Schütt et al.[21]), and then measured the performance of the model on QM9-G4MP2-holdout.

All models achieve MAEs relative to the G4MP2 atomization energy that are much lower than the MAE between B3LYP and G4MP2 atomization energies on the same molecules, 0.20 eV. We show the best-performing models in Figure 1. Our best-performing model, SchNet Delta, predicts G4MP2 energies with a MAE of only 4.5 meV (0.1 kcal/mol) after being trained on 117,232 molecules: much less than that between experiment and G4MP2 (~0.8 kcal/mol). SchNet Delta predictions are thus also an accurate estimator of experimental atomization energies.

We note interesting trends among the performance of our modified SchNet models. SchNet Delta, which uses the B3LYP energy as an input, performs best. In contrast, using the atomic partial charges as input (SchNet Charges) yields only a small performance improvement (4**%**) over baseline SchNet, most visible for smaller training set sizes. We conclude that ML models perform better when they incorporate properties that are more related to the property being predicted.

The benefits of transfer learning are also most visible on the smaller training set sizes. SchNet Transfer achieves an error of ~90 meV with only 1000 training points: 28 times better that the baseline SchNet. This result suggests a relationship between the features that best predict B3LYP and G4MP2 energies. SchNet Transfer converges faster than all other ML strategies, reaching optimal weights after only 222 epochs on our largest training set size. (Baseline SchNet requires twice as many epochs.) Re-using data clearly provides speed advantages, although the accuracy benefits decrease with training set size. The MAE of 13.4 meV for transfer learning at the largest training set size is only 10% better than baseline SchNet.

SchNet Multitask performs uniformly worse than all other models. We constructed this model by adding more outputs to the "Atomwise" output layers at the end of the network, which increases the number of parameters by less than 0.1%. This increase is apparently insufficient to simultaneously capture G4MP2 atomization energy and all other properties with the same fidelity. Given that the SchNet Transfer results indicate that the representations learned for B3LYP and G4MP2 energies are sufficiently alike to make transfer learning beneficial, the poor performance of SchNet Multitask suggests that HOMO/LUMO energies require a conflicting representation to energies or, simply, that the network requires more flexibility (e.g., more trainable weights) to model all five properties concurrently.

The performance improvements achieved by SchNet Stacked relative to baseline SchNet also diminish for large training sets. Here, the issue is that accuracy is limited by the accuracy of the underlying B3LYP model, 14 meV when trained on the full dataset; we do not exceed this accuracy in predicting the G4MP2 energies. We do note that SchNet Stacked produces accuracies comparable to SchNet Delta for small training sets and that it is superior to SchNet Transfer with the same training set size.

## 3.2 Training FCHL Models on QM9-G4MP2

We also evaluated the FCHL method, which has the best performance of any conventional machine learning strategy for predicting molecular atomization energies on the QM9 dataset to date.[14] We only tested two variants of FCHL: training directly on the atomization energies (FCHL) and training on the difference between G4MP2 and B3LYP atomization energies (FCHL Delta). As shown in Figure 1a, we find that the FCHL method achieves a MAE of 22.3 meV with a training set of $10^4$, which is consistent with the MAE reported by Faber et al. when using FHCL with B3LYP energies.[14] We were only able to train the model on training sets of $10^4$ entries or less, due to the large memory required to train, and the slow evaluation times of FCHL models with large training sets. We expect that further improvements are possible at larger training set sizes, given that the learning curve shown in Figure 1a remains roughly linear up to $10^4$ entries.

Like SchNet Delta, the FCHL Delta model also predicts G4MP2 energies very accurately, with a MAE of only 5.2 meV (0.1 kcal/mol) after being trained on $10^4$ molecules: much less than that between experiment and G4MP2 (~0.8 kcal/mol). Consequently, we do not expect that further expanding the training set of FCHL Delta will yield any improvement in the utility of the resulting models when evaluating molecules like those in QM9-G4MP2-holdout. At 5.2 meV, the error of the model is low enough that the error between G4MP2 and experiment would dominate the error between FCHL Delta and experiment. However, the fact that the learning rate for FCHL Delta has not plateaued (Figure 1a) suggests that the FCHL Delta method is flexible enough to address problems more challenging than that posed by QM9-G4MP2 (e.g., larger molecules, more diversity in chemical elements).

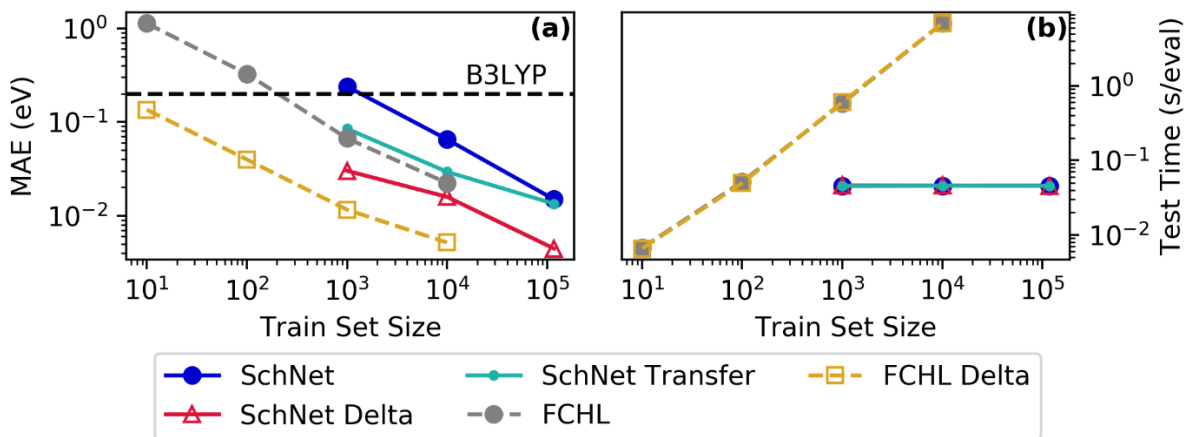

**Figure 1.** (a) Mean Absolute Error (MAE) and (b) execution speed of several ML models trained to predict the G4MP2 atomization energy. Solid lines are SchNet models,[13] and dashed lines are FCHL models.[14] Each model was trained on identical training sets of varied sizes and then tested on 13,026 total energies of molecules that were absent from the training set. The horizontal dashed line in (a) denotes the MAE of B3LYP total atomization energies compared to G4MP2 for these molecules: 0.20 eV. We limited the training set size for FCHL to $10^4$ due to its large computational cost.

### 3.3 Comparing SchNet and FCHL for Energy Prediction

We found that the FCHL models had better accuracy than the SchNet models trained with the same data. As shown in Figure 1, each FCHL model achieves accuracy comparable to that of the equivalent SchNet model (SchNet and SchNet Delta, respectively) trained with 10 times more data. The FCHL models require many fewer than 1000 G4MP2 calculations to predict the G4MP2 energies more accurately than B3LYP. SchNet Transfer performs nearly equivalently to FCHL when trained on 1000 G4MP2 calculations, illustrating the advantages of weight sharing in deep neural networks. Analogous multi-resolution training approaches in KRR[9,10] could offer a route to achieving similar improvements for the FCHL model (e.g., fewer expensive G4MP2 calculations may be necessary).

We also find a significant tradeoff between accuracy and execution speed in SchNet and FCHL. Being based on KRR, FCHL requires comparing a molecule to each molecule in its training set when predicting molecular properties, leading to an execution time that scales linearly with the number of training points (see Figure 1b). In contrast, the size of the network used in SchNet need not scale with the number of training entries and, consequently, the execution rate is invariant to training set size. FCHL achieves similar performance to SchNet with a training set of 100 entries on 10 cores of an Intel E-2680v3 CPU. Given that deep neural networks can easily use accelerators (e.g., GPGPUs) and considering that we made predictions in batches for FCHL but not for SchNet, we expect there is more room to further accelerate SchNet than FCHL.

One route to reducing the tradeoff between accuracy and execution speed is careful selection of the molecules in the FCHL training set. Browning et al. report that using a genetic algorithm to identify the best molecules can reduce KRR model error by up to 75%.[26] However, we do not expect this strategy to equalize performance, as the errors of SchNet models trained on all available data are 20 times lower than FCHL models with similar evaluation speed (i.e., those trained on 100 entries). We conclude that FCHL presents a better choice than SchNet for models of atomization energy only when training data are scarce or longer execution times are acceptable.

## 3.4 Predicting Energies of Molecules Larger than the Training Set

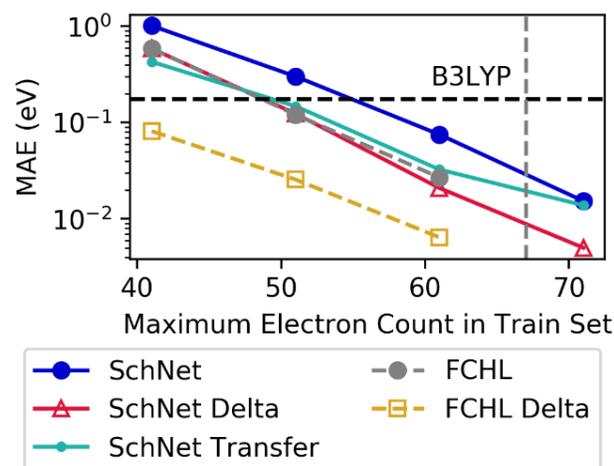

**Figure 2.** Comparison of ML models for predicting the G4MP2 atomization energies of molecules larger than those in the training set. The chart shows the MAE of models trained on various subsets of QM9-G4MP2-train, each containing all molecules that are smaller than a maximum electron count (the x axis) and evaluated using molecules larger than 67 electrons from QM9-G4MP2-holdout. The horizontal dashed line indicates the MAE of B3LYP with respect to G4MP2. The vertical dashed line marks the smallest molecule size the evaluation set.

We further tested the usefulness of our models by validating their ability to predict molecules that are larger than those in the training set. Specifically, we trained each SchNet and FCHL model with QM9-G4MP2-train subsets containing exclusively molecules with fewer than 41 (152 molecules), 51 (1161 molecules), 61 (14317 molecules), and 71 (111906 molecules) electrons, respectively. We then evaluated performance on the 6166 molecules with more than 67 electrons in QM9-G4MP2-holdout.

All of our ML models achieve more accurate predictions of atomization energy than B3LYP for large molecules, but different models require different amounts of training data to pass this threshold. As shown in Figure 2, FCHL Delta requires only the molecules with fewer than 41 electrons to improve upon B3LYP. In contrast, the best SchNet models and the baseline FCHL model require all molecules with fewer than 51 electrons to predict atomization energies better than B3LYP. All models outperform B3LYP when they are trained on all molecules smaller than 61 electrons. At that training set size, SchNet Delta and FCHL Delta predict G4MP2 atomization energies with a MAE 8.5 and 27.7 times lower than that of B3LYP, respectively, which illustrates how our ML models can generate accurate predictions for molecules larger than those in the training set.

We also studied the effect of molecule size on model error, with the goal of understanding how our models are likely to perform for molecules even larger than those in QM9-G4MP2. We trained each model on all molecules in QM9-G4MP2-train with fewer than 61 electrons and measured model performance for subsets of molecules from QM9-G4MP2-holdout with different sizes. The MAEs for each model generally increase with molecule size yet are all lower than the MAE of B3LYP. This increase in error with molecule size cannot be explained completely by the energy scale of the molecules increasing with size. As shown in Figure 3, we find that the error in energy per electron also tends to increase with molecular size. Given that the difference between B3LYP and G4MP2 energies remains roughly constant with molecule size (Figure 3), we expect that there is a maximum molecule size beyond which our models will fail to predict GM4MP2 energies accurately. However, we lack sufficient data to estimate the cross-over point with confidence.

To further test the ability of our models to predict the energies of large molecules, we compared model predictions to the G4MP2 results in the G4MP2-heavy data set (see Section 2.1). We first trained SchNet Delta on all 117,232 entries in QM9-G4MP2-train and FCHL Delta on the largest training set feasible on our hardware, $10^4$ entries. We found that the FHCL Delta and SchNet Delta models achieve MAEs of 12.5 meV and 39.5 meV, respectively, on G4MP2-heavy. The results are shown in Figure 5. These errors are somewhat higher than those measured on QM9-G4MP2-holdout (4.4 meV and 5.1 meV, respectively), which is consistent with our finding that errors increase with molecule size. That said, the models predict G4MP2 energies with substantially more accuracy than the B3LYP energies for these molecules (MAE 439.1 meV). FCHL and SchNet Transfer, which do not use the B3LYP energy as input, achieve MAEs that are better estimates of the G4MP2 energy than B3LYP but are worse than the ∆-learning models—of 52.2 and 110.8 meV, respectively. The 10 times or greater improvement in the atomization energy of the ∆-learning models further supports our conclusion that our models can be used for larger molecules. For now, we estimate that the models are applicable to molecules with at least 14 heavy atoms but recommend further studies, due to the small test set size of 66 molecules.

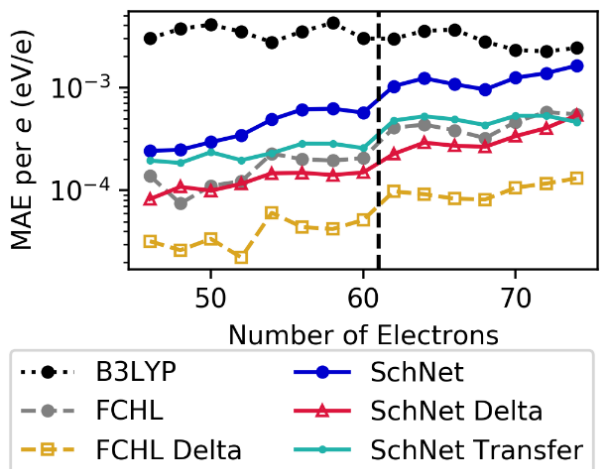

**Figure 3.** Effect of molecule size on prediction error. The chart shows, for subsets of the test data with different electron counts, the MAE normalized by the number of electrons between five ML models and G4MP2. Models were trained only on molecules with fewer than 61 electrons from QM9-G4MP2-train and evaluated against molecules from QM9-G4MP2-heavy; a vertical dashed line at 61 electrons is provided for clarity. We also show, as B3LYP, the equivalent error metric between B3LYP and G4MP2 energies.

### 3.5 Sensitivity of Predictions to Molecular Coordinates

A key limitation of SchNet and FCHL is that, in contrast to molecular-graph-based methods (e.g., Refs. [22,27–31]), both require 3D coordinates to predict molecular properties. As each model was trained using the B3LYP equilibrium structures as input, we assume that it may be necessary to first determine the equilibrium structure with B3LYP before predicting atomization energy—unless a suitable approximation is available. Removing the need to perform a B3LYP calculation to determine the equilibrium structure would drastically accelerate the rate at which we can predict G4MP2 energies. Consequently, we studied whether energies can be computed accurately when using, instead, atomic coordinates generated by using cheap force fields.

Our first step was to study the effect of different methods for guessing atomic coordinates on model accuracies. We used Open Babel to approximate atomic coordinates algorithmically using known bond angles, by relaxing the structure using the MMFF94 force field,[32] and using a search for the lowest energy conformation among different permutations of rotatable bonds.[33] We used the resulting coordinates as input to the G4MP2 Transfer model trained on all of QM9-G4MP2-train, and then determined the MAE of the model with respect to QM9-G4MP2-holdout. The model evaluated using the structure post conformer search achieved the lowest MAE of 327 meV—higher than the B3LYP atomization energy MAE of 201 meV and 25 times larger than the error when using the B3LYP coordinates. We found similar performance degradation for SchNet Stacked. We conclude that the models retain some predictive power (a MAE of 327 meV is 200 times better than a guess-the-mean model) when used on approximate coordinates, but that performance is degraded enough to make the models significantly less useful.

We attempted two different transfer learning routes for improving the performance of models on approximate coordinates. First, we retrained SchNet Transfer with the approximate coordinates as input and using the weights learned using all of QM9-G4MP2-train as a starting point; this reduces the MAE to 222 meV. We further reduced the dependence on coordinates by selecting a different conformer for each molecule at each epoch and rattling the coordinates of the conformer with a standard deviation of 0.1 Å. Retraining with "blurred" coordinates of molecules reduced the MAE to 205 meV, which is equivalent to the B3LYP atomization energy for the same molecules (200 meV). In short, our model produces estimates of the G4MP2 atomization energy for molecules with nine or fewer heavy atoms that are as good as B3LYP, but at much faster rates.

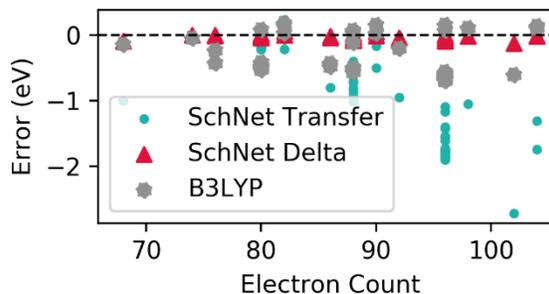

**Figure 4.** Accuracy of a (cyan) SchNet Transfer model retrained on the conformers each molecule in QM9-G4MP2-train dataset, the SchNet Delta model (red), and (gray) B3LYP on predicting the G4MP2 atomization energy. The SchNet Transfer was trained by selecting a random conformer for each molecule and then coordinates were perturbed with a standard deviation of 0.1 Å at each epoch. Each point represents the error between the prediction and G4MP2 for a molecule with more heavy atoms than those in the training set.

Unfortunately, the strategy of re-training the model on perturbed coordinates fails for larger molecules. As shown in Figure 4, the baseline SchNet model retrained on the perturbed conformers for each molecule has errors of up to 2 eV when predicting the energy of a molecule given its generated atomic coordinates. The errors increase with molecular size, which could be both an effect of the error of ML models increasing with molecular size and the quality of the generated coordinates decreasing. The RMSD between the generated and B3LYP coordinates increases with the number of heavy atoms in our large model test set; for the largest molecules (14 heavy atoms), it is over four times larger than the median RMSD in our small molecule test set.

We conclude that it is possible to reduce the dependence of SchNet-based models on knowing the DFT-relaxed coordinates of a molecule by training on the conformers of each molecule with perturbed coordinates. The retrained models can predict the G4MP2 energies of small molecules with superior accuracy to B3LYP when provided only the coordinates generated with force fields. However, this strategy currently fails for larger molecules. Considering the higher accuracy of our models when given the B3LYP coordinates (Section 3.4), better accuracy on large molecules could be achieved with better estimates of relaxed coordinates (e.g., by using generative networks[34]). However, we do not currently recommend using SchNet to predict atomization energies without relaxed coordinates until improved techniques for generating atomic coordinates are available.

## 3.6 Recommendations

Our results lead us to recommend two different ML models for predicting accurate energetics at the G4MP2 level of theory, depending on performance needs. If optimized B3LYP coordinates and energies are available, we recommend using the FCHL Δ-learning model for optimal accuracy and the SchNet Δ-learning model if slower time-per-prediction of the FCHL model is unacceptable. As shown in Figure 5, both models can be used to increase the accuracy of energies relative to B3LYP calculations by up to a factor of 10.

The SchNet Delta and FCHL Delta models are available for anyone to use via DLHub.[15] DLHub's simple Python interface takes the XYZ coordinates of a molecule and returns the G4MP2 atomization enthalpy; it runs models on cloud or cluster resources, eliminating the need to understand how to use QML or SchNetPack or even to install them. We hope that by publishing the models in this way, we will enable others to integrate the capabilities developed in this work in their own research.

## 4 Conclusion

We compared multiple ML strategies for producing accurate estimates of G4MP2-level atomization energies of molecules at reduced computational costs. We evaluated models based on the FCHL and SchNet approaches and found that both approaches yield models that reliably predict the atomization energy 10 times more accurately than B3LYP for molecules larger than those in their training sets. The strong performance was achieved by learning the difference between B3LYP and G4MP2 atomization energies—an approach that we found yields higher accuracy than other methods of integrating data from multiple fidelities of calculations (e.g., transfer learning). We produced two state-of-the-art ML models that predict the G4MP2-level atomization energy of molecules within 5 meV for molecules of similar size and somewhat larger for molecules larger than those in the training set. Our FCHL-based model has accuracies three times better than SchNet but has at least a 100 times greater computational cost per prediction. We have made the models available for unrestricted use via a web API.

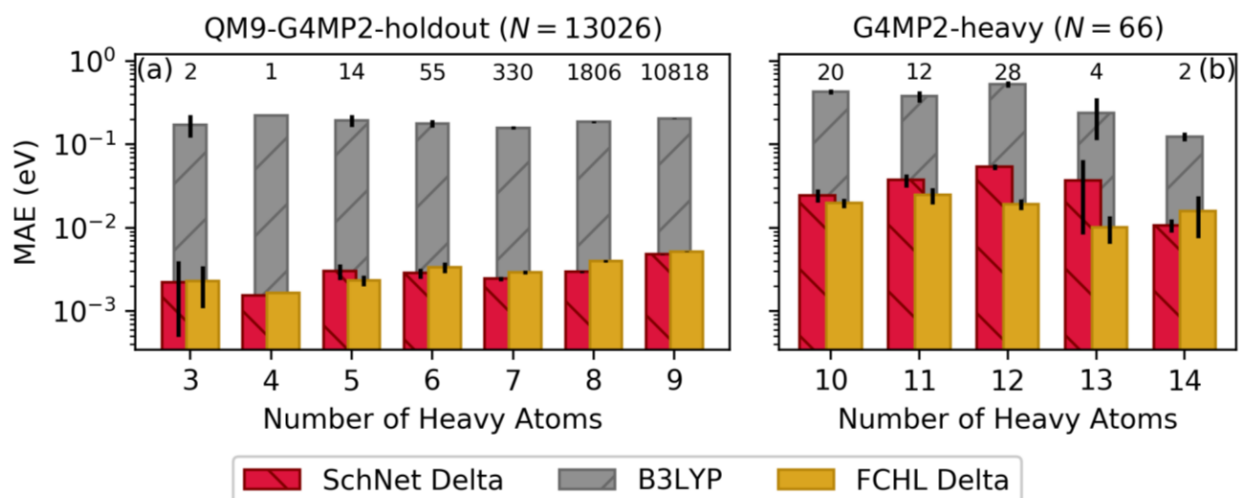

**Figure 5.** MAEs between G4MP2-level atomization energy calculations and SchNet Delta predictions, FCHL Delta predictions, and B3LYP-level energy calculations. The two ML models were trained, using only molecules with nine heavy atoms or fewer, to predict the difference between B3LYP and G4MP2 energies. MAEs are shown relative to two sets of molecules, both outside of the ML models' training set: (a) a holdout set of molecules with nine or fewer heavy atoms, and (b) a small dataset of molecules with 10 or more heavy atoms. The error bars are the standard error of the mean. Numbers above each bar indicate the number of molecules with each number of heavy atoms.

## 5 Acknowledgements

This research was supported in part by the Exascale Computing Project (17-SC-20-SC) of the U.S. Department of Energy (DOE), by DOE's Advanced Scientific Research Office (ASCR) under contract DE-AC02-06CH11357, and by the Joint Center for Energy Storage Research (JCESR), an Energy Innovation Hub funded by the U.S. Department of Energy, Office of Science, Basic Energy Sciences. This work used resources from the Extreme Science and Engineering Discovery Environment (XSEDE), supported by National Science Foundation Grant No. ACI-1548562:[35] specifically, Jetstream at the Texas Advanced Computing Center through allocation CIE170012;[36] the University of Chicago Research Computing Center; and the Argonne Leadership Computing Facility. This material is based upon work supported by Laboratory Directed Research and Development (LDRD) funding from Argonne National Laboratory, provided by the Director, Office of Science, of the U.S. Department of Energy under Contract No. DE-AC02-06CH11357. This work was performed under financial assistance award 70NANB14H012 from U.S. Department of Commerce, National Institute of Standards and Technology as part of the Center for Hierarchical Material Design (CHiMaD). This work was also supported by the National Science Foundation as part of the Midwest Big Data Hub under NSF Award Number: 1636950 "BD Spokes: SPOKE: MIDWEST: Collaborative: Integrative Materials Design (IMaD): Leverage, Innovate, and Disseminate".